\documentclass[%
reprint,
twocolumn,
amsmath,amssymb,
aps,
prb,
]{revtex4}

\usepackage{graphicx}
\usepackage{dcolumn}
\usepackage{bm}
\usepackage{color}

\begin{document}

\title{First-Principles investigation of the First-order Phase Transition in Kagome $\mathbf{Na_2Ti_3Cl_8}$}

\author{Shuang Liu}
\author{Zhi-Bo Yin}
\author{Meng-Qiu Long}
\author{Yun-Peng Wang}
\email{yunpengwang@csu.edu.cn}
\affiliation{%
School of Physics and Electronics,
Hunan Key Laboratory for Super-micro structure and Ultrafast Process,
Central South University, 932 South Lushan Road, Changsha, P. R. China
}%

\date{\today}

\begin{abstract}
The antiferromagnetic kagome material $\mathrm{Na_2Ti_3Cl_8}$ exhibits a first-order phase transition;
the low-temperature phase is characterized by the trimerizations of Ti ions.
In this work, we carry out first-principles calculations on the crystal and electronic structure
of $\mathrm{Na_2Ti_3Cl_8}$ at the high- and low-temperature phases.
The crystal structures, including the lattice constant and the trimerization of Ti ions,
are well reproduced by taking account a small Hubbard correction of $1 \, \mathrm{eV}$.
The calculated total energy landscape reveals a first-order phase transition
with the total energy barrier of $80 \,\mathrm{meV}$ per formula unit.
Analysis of the electronic structure indicates a direct metal bonding among Ti
in the trimierized low-temperature phase.
\end{abstract}

\maketitle

\section{Introduction}

The antiferromagnetic kagome $\mathrm{Na_2Ti_3Cl_8}$ material undergoes
a nonpolar-to-polar first-order phase transition as the temperature drops to around 200\,K.
\cite{Na2Ti3Cl8-beta-phase,Na2Ti3Cl8-polar}
The Ti ions form a perfect kagome lattice in the high-temperature (HT) phase but trimerize in the low-temperature (LT) phase.
The crystal symmetry is also reduced.
The space group of the HT phase is $R\bar{3}m$ but is $R3m$ in the LT phase.
The phase transition is also characterized by a numerous reduction of the in-plane lattice constant
(from HT $7.4\,\mathrm{\AA}$ to LT $7.0\,\mathrm{\AA}$),
and a substantial reduction ($\sim 9 \%$) in the volume from 0.315 to 0.287 $\mathrm{nm}^3$ per formula unit.
The thermal hysteresis of the phase transition is about 25\,K.\cite{Na2Ti3Cl8-beta-phase}
Both the LT and HT phases are antiferromagnetic,
and a high-spin to low-spin transition occurs along with the structure transition.\cite{Na2Ti3Cl8-1995}
During the cooling process, an intermediate ordered phase emerges
with only half of Ti ions participating in the trimerization.\cite{Na2Ti3Cl8-beta-phase}

The exotic antiferromagnetic order on the kagome lattice is interesting by itself,
and was used as the starting point for understanding the phase transition in $\mathrm{Na_2Ti_3Cl_8}$.\cite{Na2Ti3Cl8-spin-lattice}
A trimerized ground state emerges spontaneously from a Heisenberg antiferromagnet on kagome lattice.\cite{trimer-AFM-kagome}
The trimerized magnetic order coincides with the Ti ion trimerization in the $\mathrm{Na_2Ti_3Cl_8}$ material.
In a recent work\cite{Na2Ti3Cl8-spin-lattice},
the phase transition in $\mathrm{Na_2Ti_3Cl_8}$ was interpreted as
driven by the coupling between the spin and the lattice degrees of freedom.
The rotational symmetry was proposed to be spontaneously broken by the trimerized magnetic state,
which acts as a precursor driving the structural phase transition.\cite{Na2Ti3Cl8-spin-lattice}
Recalling that experimental data unambiguously indicate that the phase transition in $\mathrm{Na_2Ti_3Cl_8}$ is of first-order.
A first-order phase transition, in general, does not require a spontaneous instability in the high-temperature phase
as a guiding and driving force towards the LT phase.
Instead, a lower free energy of the LT phase is sufficient for guarantee the phase transition.
Theoretical understanding of the first-order phase transition and the changes of electronic structure after the phase transition is still lacking.

\begin{figure}[h!]
\centering
\includegraphics[width=\linewidth]{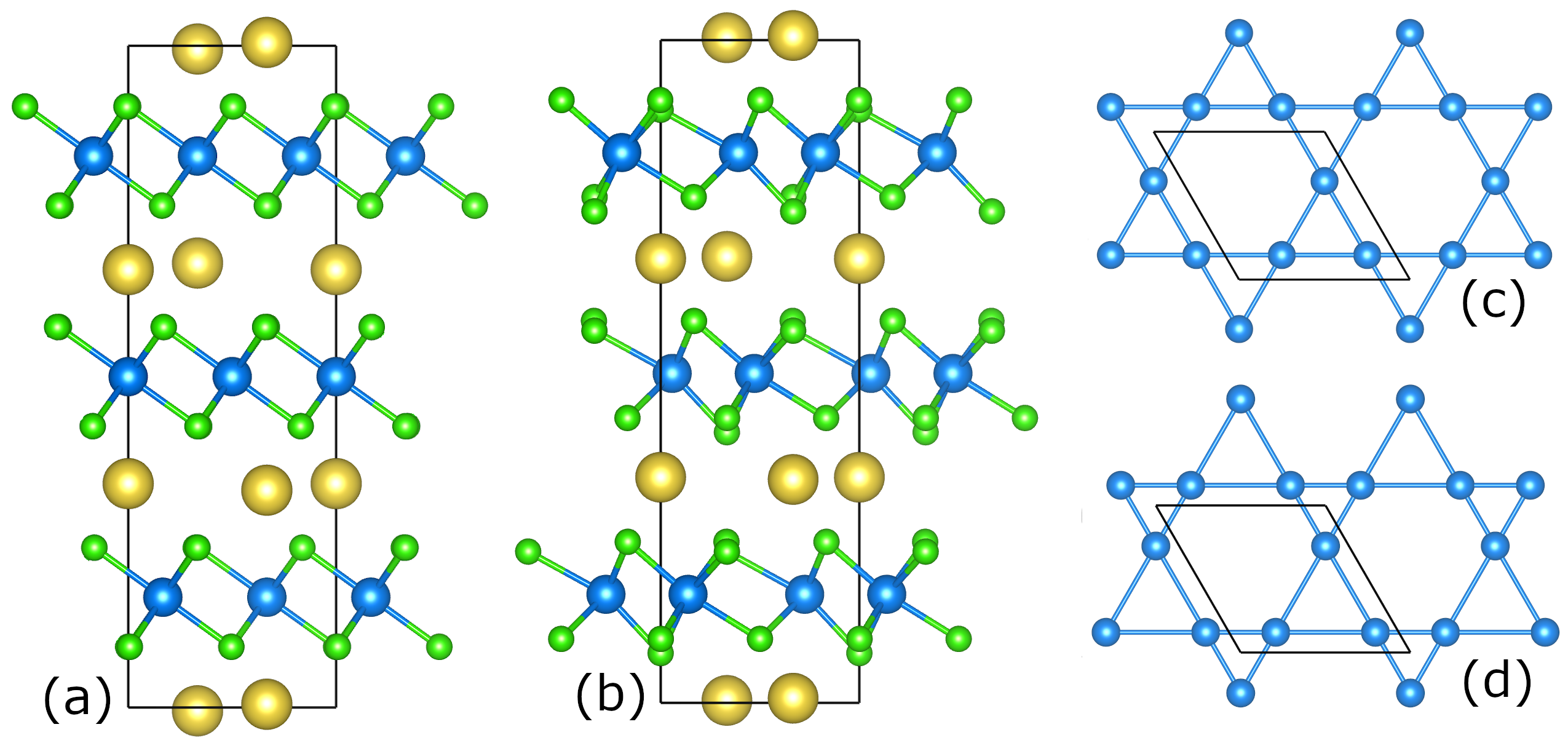}
\caption{
\label{fig:structure}
The crystal structure of the unit cell of $\mathrm{Na_2Ti_3Cl_8}$ in (a) the high-temperature phase and the (b) the low-temperature phase.
The (c) perfect and (d) distorted with trimerizations of the Ti kagome lattice for the high- and low-temperature phases respectively.
}
\end{figure}

In this work, we provide a first-principles based description of the HT and LT phases of $\mathrm{Na_2Ti_3Cl_8}$ and the first-order phase transition between them.
The calculated lattice structures, including the lattice constant and the trimerization of Ti ions, agree with experimental data.
The zero-temperature energy landscape possesses an energy barrier of $0.08 \, \mathrm{eV}$ per formula unit from the HT phase to the LT phases.
Electronic structure analysis reveal the splitting of Ti-$d$ states
and a direct metal-metal bonding in the LT phase.

\section{Computational details}

Density functional theory (DFT) calculations are conducted
using the Vienna Ab initio Simulation Package(VASP)\cite{vasp1,vasp2,vasp3,vasp4,vasp5}.
We use the semilocal generalized gradient approximation (GGA) of Perdew-Burke-Ernzerhof (PBE)\cite{PBE}.
The intra-atomic electronic correlation among $d$-orbitals of Ti ions were taken into account
using the DFT+U method\cite{DFTU} with the Hubbard $U$ parameter set to between 0 and 4~ eV.
The plane-wave cutoff energy is set to $400\,\mathrm{eV}$ in all the calculations.
The first Brillouin zone is sampled using a uniform $3 \times 3 \times 1$ $k$-point mesh.
The crystal structure was optimized until the forces on each atom are smaller than $0.001\,\mathrm{eV/\AA}$.
The solid-state climbing-image nudged elastic band calculations are performed using
the VASP Transition State Tools (VTST).\cite{NEB2000,CINEB2000,SSNEB2012}

\section{Results and Discussions}

\subsection{Crystal structures}


The $\mathrm{Na_2Ti_3Cl_8}$ material has a layered structure as depicted in Fig.~\ref{fig:structure}.
The side views show that covalently bonded Ti-Cl layers separated by Na ions, see Fig.~\ref{fig:structure}(a).
For each Ti-Cl layer, one Ti cation is located within an octahedron formed by six Cl anions.
The Cl octahedra are perfect and the Ti ion is at its center for the HT phase,
while the Cl octahedra strongly distort in the LT phase.
The Ti ions of each Ti-Cl layer form a regular kagome lattice at the HT phase, as shown in Fig.~\ref{fig:structure}(c).
However, at the LT phase the kagome lattice distorts to make Ti ions no longer equal-distanced, see Fig.~\ref{fig:structure}(d).

\begin{figure}[h!]
\centering
\includegraphics[width=\linewidth]{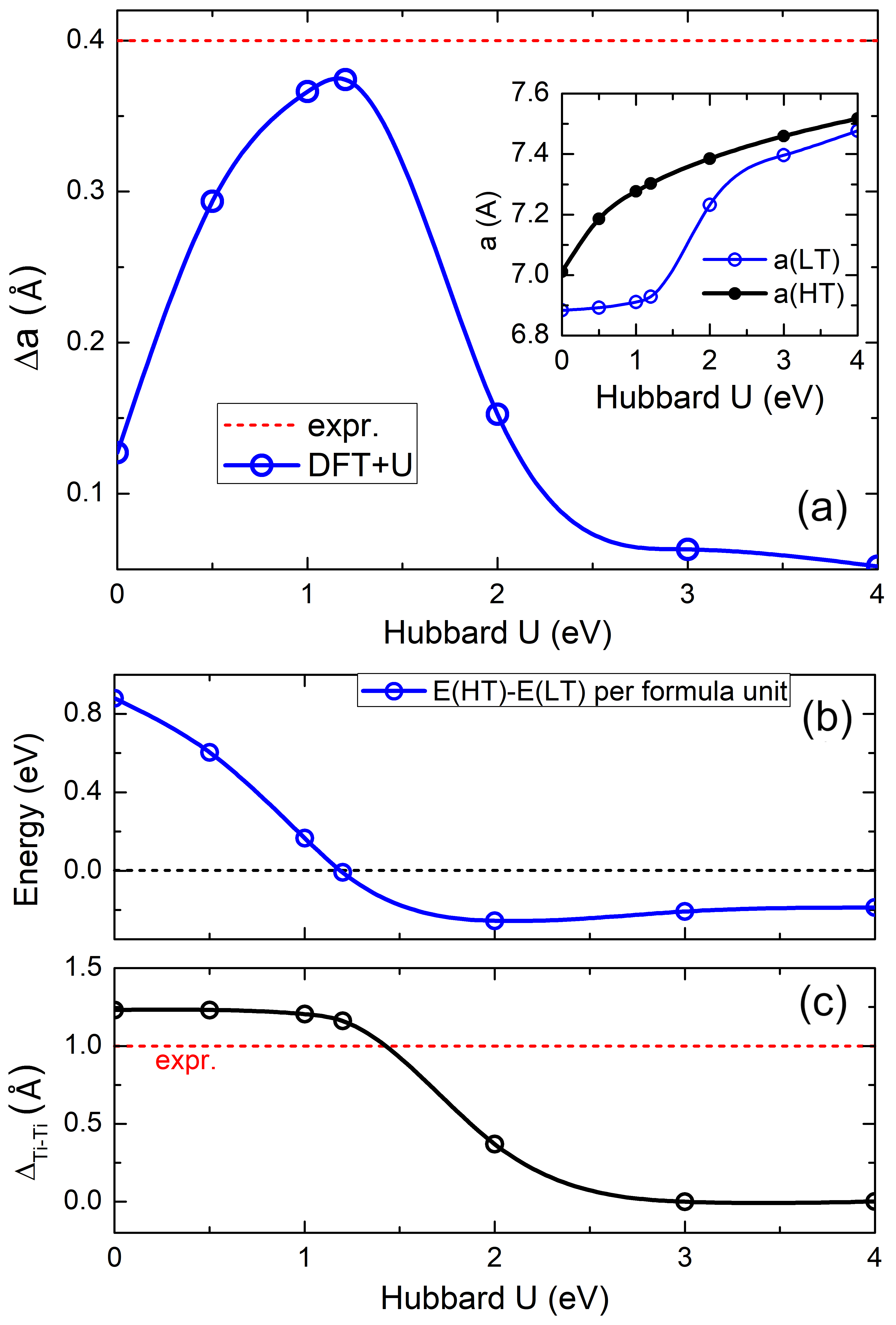}
\caption{
\label{fig:PBEU}
The Hubbard U dependence of (a) the lattice constant difference,
(b) the total energy difference between the low-temperature and high-temperature phases,
and (c) the difference between side lengths of Ti-triangles in the low-temperature phase.
}
\end{figure}

We first study the crystal structures of $\mathrm{Na_2Ti_3Cl_8}$ in the LT and HT phases.
Both the lattice vectors and the atomic coordinates within the unit cell are allowed to relax.
Structural optimizations starts from the experimental LT and HT structures\cite{Na2Ti3Cl8-beta-phase}.
The initial magnetic configuration is set to
the collinear ferromagnetic or the $120^\circ$ noncollinear antiferromagnetic configurations.
Experimental data suggest an antiferromagnetic ground state.
Applications of the DFT to materials containing transition-metal elements
usually requires an additional and empirical correction of the electron-electron Coulomb repulsion,
which is not sufficiently accounted for in the local and semilocal density functionals.
As an early transition metal, the Ti element usually requires only a small Hubbard correction.
We therefore perform a series of DFT calculations
with the Hubbard $U$ parameter set to a value between $0.0 \,\mathrm{eV}$ and $4.0 \, \mathrm{eV}$.

The calculation results are shown in Fig.~2.
The $120^\circ$ noncollinear antiferromagnetic configuration is almost always the ground state;
the lattice constant of the HT phase is always longer than the LT phase;
both are in accord with experiments.
We look into three key physical quantities and compare the calculated values with the experimental data.
The first one is the difference of lattice constant $a$ of the LT and HT phases,
defined as $\Delta a = a(HT) - a(LT)$.
We omit the lattice constant $c$ because the trimerization of Ti ions happens within the $ab$-plane.
For the cases with Hubbard $U$ set to be larger than $2.0 \, \mathrm{eV}$,
the calculated lattice constant of the HT phase is only slightly longer ($\sim 0.05 \, \mathrm{\AA}$) than the LT phase,
which may be caused by numerical inaccuracy; in fact,
the LT structure relaxes to the HT structure for these cases.
Usage of a such larger $U$ shall be avoided.
For the Hubbard $U$ increases from $0.0$ to $2.0 \,\mathrm{eV}$,
the calculated $\Delta a$ increases and then decreases.
At $U=1\,\mathrm{eV}$, the calculated $\Delta a$ reaches its maximum of $0.38 \, \mathrm{\AA}$
(experimental data: $\sim 0.4 \,\mathrm{\AA}$)\cite{Na2Ti3Cl8-1995,Na2Ti3Cl8-beta-phase,Na2Ti3Cl8-polar}.
The calculated lattice constant $a$ of LT and HT phases matches experimental values in parentheses:
$7.32\,\mathrm{\AA}$ ($7.43\,\mathrm{\AA}$) for the HT phase and
$6.93\,\mathrm{\AA}$ ($6.98\,\mathrm{\AA}$) for the LT phase.\cite{Na2Ti3Cl8-1995}

The second key quantity characterizes the trimerization of Ti cations.
In the HT phase, the Ti ions form a perfect kagome lattice,
which comprises of two types of Ti-triangles with opposite orientations.
In the LT phase, the side lengths of one type of Ti-triangles decrease,
while the other type of Ti-triangles expand,
resulting in the so-called Ti trimerization.
The difference between side lengths of Ti-triangles (denoted as $\Delta_\mathrm{Ti-Ti}$)
is used for characterizing the amplitude of the Ti trimerization.
The experimental value of Ti-Ti distances are $3.071 \,\mathrm{\AA}$ and $3.903 \,\mathrm{\AA}$ for the LT phase,
thus experimental $\Delta_\mathrm{Ti-Ti}$ is about $0.9\,\mathrm{\AA}$.\cite{Na2Ti3Cl8-beta-phase}
Fig.~2(b) shows that the calculated $\Delta_\mathrm{Ti-Ti}$ is around $1.2 \,\mathrm{\AA}$
when Hubbard $U$ is lower than $1 \,\mathrm{eV}$.
For $U$=$2 \,\mathrm{eV}$ and larger, the calculated $\Delta_\mathrm{Ti-Ti}$ vanishes.
Therefore the PBE+U calculations with $U > 2 \,\mathrm{eV}$ fails to predict the Ti trimerization in the LT phase.

The third quantity is the total energy difference: $\Delta E=E_{\mathrm{HT}} - E_{\mathrm{LT}}$,
which shall be positive to guarantee the LT phase as the ground state.
This criteria is satisfied when the Hubbard $U$ is smaller than $1 \,\mathrm{eV}$.
As the Hubbard $U$ decreases further, the $\Delta E$ increases to $0.9 \,\mathrm{eV}$ per formula unit at $U=0 \,\mathrm{eV}$.
The value of $\Delta E$ shall not be too much, otherwise the HT phase is not possible to appear.
This consideration forces us to choose the Hubbard $U$ to be around $1\,\mathrm{eV}$.
Overall, according to the behaviors of the aforementioned three quantities versus the Hubbard $U$,
we choose $U=1\,\mathrm{eV}$, which reproduces all the experimental observations fairly well.

\subsection{The first-order phase transition}

Next, we confirm that the phase transition is of first-order by calculating the total energy landscape.
The Wyckoff positions of Ti atoms in the HT phase are $(0.5, 0.5, 0)$
but become $(0.5+\delta_{\mathrm{Ti}}, 0.5-\delta_{\mathrm{Ti}},0)$ in the LT phase.
The distortion of Ti kagome lattice is characterized by $\delta_{\mathrm{Ti}}$,
which is one of the order parameter for the phase transition:
$\delta_{\mathrm{Ti}} = 0$ for the HT phase and $\delta_{\mathrm{Ti}} = 0.02 $ for the LT phase,
according to experimental data.\cite{Na2Ti3Cl8-beta-phase}
Other quantities deeply involved in the phase transition is the lattice constants $a$ and $c$;
both exhibit significant changes and hysteresis during the phase transition.\cite{Na2Ti3Cl8-beta-phase}
The quasi-two-dimensional nature of the $\mathrm{Na_2Ti_3Cl_8}$ material
makes it natural to choose the lattice constant $a$, but not $c$ as the other order parameter.
We construct the total energy landscape in the order parameter phase spanned by $\delta_{\mathrm{Ti}}$ and $a$.
In practice, we optimize the lattice constant $c$ and positions of all atoms except Ti,
while the lattice constant $a$ and the fractal positions of Ti ions are fixed.

\begin{figure}[h!]
\centering
\includegraphics[width=\linewidth]{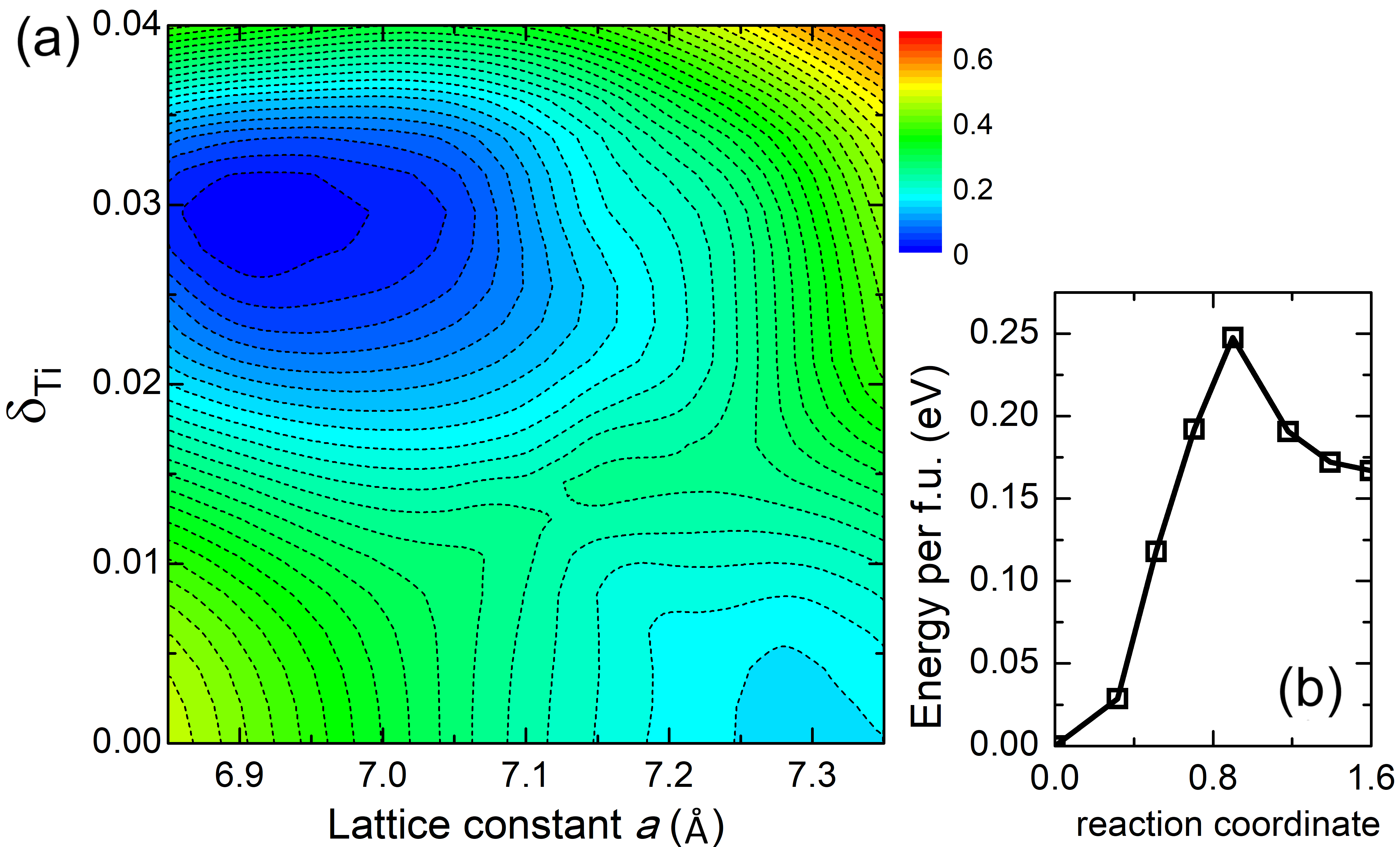}
\caption{
\label{fig:landscape}
(a) The calculated total energy landscape in the $a$-$\delta_\mathrm{{Ti}}$ plane,
where $a$ is the lattice constant in unit of $\mathrm{\AA}$ and $\delta_\mathrm{{Ti}}$ is the deviation of Wyckoff positions (see main text).
(b) The calculated minimum energy path between the low- and high-temperature phases.
The total energy of the low-temperature phase is set to zero as a reference.
}
\end{figure}

The calculated total energy landscape is shown in Fig.~\ref{fig:landscape}.
The LT phase corresponds to $\delta_{\mathrm{Ti}} = 0.029$ and $a=6.91 \,\mathrm{\AA}$ in Fig.~\ref{fig:landscape},
while the HT phase corresponds to $\delta_{\mathrm{Ti}}=0.0$ and $a=7.28 \,\mathrm{\AA}$.
The total energy landscape in Fig.~\ref{fig:landscape} confirms that
both the LT and the HT phases are located at the local minima of the energy landscape,
hence the transition between them shall be of first-order.
One can read an energy barrier of about $0.8 \, \mathrm{eV}$.
We also conduct a solid-state climbing-image nudged elastic band\cite{SSNEB2012} calculation on the transition from the LT to the HT phase.
The total energy relative to that of the LT phase along the reaction coordinate is plotted in Fig.~\ref{fig:landscape}(b).
The transition state, denoted by the star in Fig.~\ref{fig:landscape}(b) is higher by $0.74 \, \mathrm{eV}$ than the LT phase,
and $0.24 \, \mathrm{eV}$ than the HT phase.

\subsection{Electronic structures}

The Ti ions in $\mathrm{Na_2Ti_3Cl_8}$ have a nominal valence of +2, thus in the $d^2$ configuration.
The crystal field of Ti-$d$ electrons due to Cl ions in the first shell is perfectly octahedral in the HT phase,
hence the Ti-$d$ states split to triply degenerate $t_{\mathrm{2g}}$ orbitals but only occupied by two $d$ electrons.
Therefore the Ti ions exhibit unquenched orbital moments.
The $\mathrm{V^{2+}}$ ions in the $\mathrm{VI_3}$ material exhibit unqueched orbital moments in a similar manner.
However the orbital moments of $\mathrm{Ti^{2+}}$ ions in $\mathrm{Na_2Ti_3Cl_8}$ is smaller.
The spin-only magnetic moment is $g\sqrt{S(S+1)}\mu_B = 2.83 \,\mu_B$ (setting the $g$-factor to 2.0),
which is only slightly larger than experimental value of $2.55 \, \mu_B$ at $300\,\mathrm{K}$.
This indicates nearly quenched orbital moments for $\mathrm{Ti}^{2+}$ ions in the HT phase.
Our DFT calculations including the spin-orbit coupling effect confirm that
the orbital magnetic moment of $\mathrm{Ti}^{2+}$ ions is smaller than $0.06 \, \mu_B$.
For the LT phase the calculated orbital magnetic moments is about $0.01 \, \mu_B$ per Ti ion
although the Cl-octahedra are strongly distorted.
The atomic magnetic moments thus are dominated by the spin moment,
which is $1.51 \,\mu_B$ and $0.53 \,\mu_B$ per Ti ion for the HT and the LT phases according to our DFT calculations.
Although the DFT underestimates the magnetic moments by about $1.0 \, \mu_B$
(experimental values are $2.55 \, \mu_B$ and $1.49 \, \mu_B$ for HT and LT phases),
DFT reproduces the $ \sim 1.0 \, \mu_B$ reduction of magnetic moments from the HT to the LT phase.

\begin{figure}[h!]
\centering
\includegraphics[width=\linewidth]{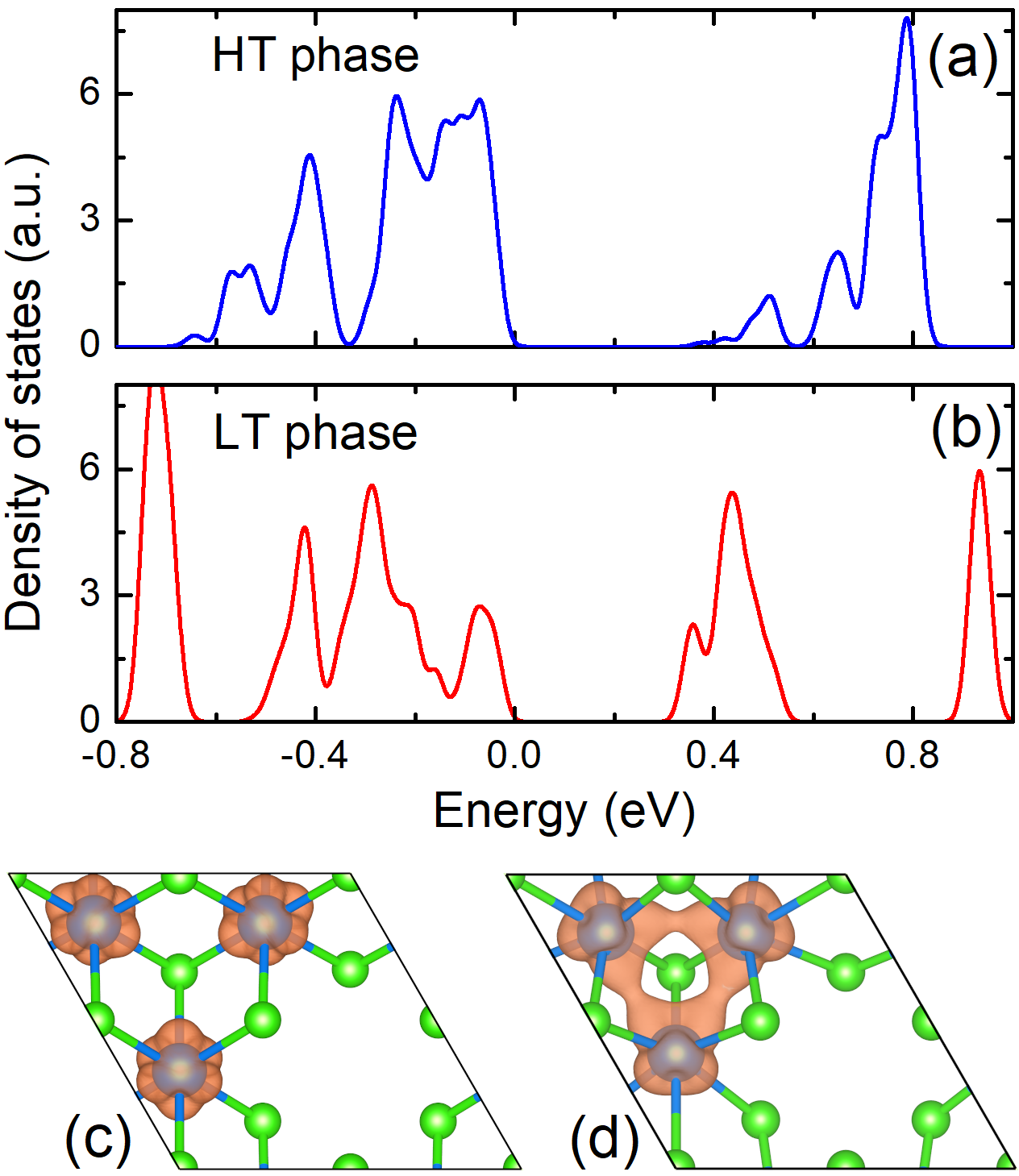}
\caption{
\label{fig:DOS}
The calculated electronic density of states of (a) the high-temperature and (b) the low-temperature phases.
The real-space distribution of the charge density corresponding to the occupied states shown in (a,b)
for (c) the HT and (d) the LT phases;
the iso-value of the charge density is set to $0.06 / \mathrm{\AA}^{3}$.
Color code for species: green for Cl and blue for Ti; Na ions are omitted for clarity.
}
\end{figure}

The $\mathrm{Na_2Ti_3Cl_8}$ material is semiconducting and the calculated band gap is
about $0.3 \, \mathrm{eV}$ for both the HT and the LT phases, as shown in Fig.~\ref{fig:DOS}(a,b).
The application of intra-atomic Coulomb repulsion makes the triply degenerate $t_{\mathrm{2g}}$ orbitals split.
For the HT phase, the unoccupied DOS has a peak at $0.8 \, \mathrm{eV}$ and the occupied DOS has a peak at $-0.2 \, \mathrm{eV}$;
their difference of $1.0 \, \mathrm{eV}$ comes from the Hubbard $U$ of the same value.
The additional trimerization distortions in the LT phase split the DOS further, as indicated by Fig.~\ref{fig:DOS}(b).
A strong DOS peak is detached and resides at $-0.7 \,\mathrm{eV}$, lower than other valence states.
The unoccupied DOS also shows a clear splitting of about $0.5 \,\mathrm{eV}$.

\section{Conclusions}

In summary, we re-investigated the phase transition in kagome material $\mathrm{Na_2Ti_3Cl_8}$.
We confirm the first-order nature of the phase transition as observed in experiments,
and we predict an energy barrier of 0.24\,eV per formula unit for this transition.
The changes in the lattice constant and the local magnetic moments are also in accord with experimental data.
The trimerization of Ti ions triggers a direct Ti-Ti bonding.

\begin{acknowledgments}
This work is supported by the National Natural Science Foundation of China (Grant No. 12004439)
and computational resources from the High Performance Computing Center of Central South University.
\end{acknowledgments}


\end{document}